\documentclass[12pt]{iopart}

\pdfoutput=1

\usepackage{fancyhdr}

\usepackage{bm}

\usepackage[english]{babel}

 \expandafter\let\csname equation*\endcsname\relax

 \expandafter\let\csname endequation*\endcsname\relax

\usepackage{amsmath,amssymb,amsfonts,latexsym}

\usepackage{graphicx}

\usepackage{color}

\usepackage{iopams}

\usepackage{afterpage}

\usepackage{cite}

\usepackage{tikz}

\usetikzlibrary{snakes}

\usepackage{xcolor}

\usepackage{float}

\usepackage{times}

\usepackage[colorlinks,citecolor=blue,linkcolor=red,urlcolor=blue]{hyperref}

\newcommand{\be}{\begin{equation}}

\newcommand{\ee}{\end{equation}}

\newcommand{\bea}{\begin{eqnarray}}

\newcommand{\eea}{\end{eqnarray}}

\usepackage{etoolbox}

\makeatletter

\def\@mkboth#1#2{}

\newlength\appendixwidth

\preto\appendix{\addtocontents{toc}{\protect\patchl@section}}

\newcommand{\patchl@section}{%

  \settowidth{\appendixwidth}{\textbf{Appendix }}%

  \addtolength{\appendixwidth}{1.5em}%

  \patchcmd{\l@section}{1.5em}{\appendixwidth}{}{\ddt}%

}

\begin{document}

\title[]{Scaling behavior of Ising systems at first-order transitions}

\author{Pierpaolo Fontana}

\address{SISSA, via Bonomea 265, 34136 Trieste, Italy}
\ead{pfontana@sissa.it}
\vspace{10pt}
\begin{indented}
\item[]\date{\today}
\end{indented}

\begin{abstract}
We investigate how the scaling behavior of finite systems at magnetic first-order transitions (FOTs) with relaxational dynamics changes in correspondence of various boundary conditions. As a theoretical laboratory we consider the two-dimensional Ising model in the low-temperature phase. When the boundary conditions do not favor any specific phase of the system, we show that a dynamic finite-size scaling (DFSS) theory can be developed to describe the dynamic behavior in the coexistence region, where different phases coexist. When the boundary conditions at two opposite sides of the system generate a planar interface separating the phases, we show that the autocorrelation times are characterized by a power-law behavior, related to the dynamics enforced by the interface. Numerical results for a purely relaxational dynamics confirm the general picture.
\end{abstract}

\section{\label{intro}Introduction}
Close to a phase transition point, thermodynamic functions develop a singular behavior in the thermodynamic limit, i.e. when the volume of the considered system tends to infinite.  However, in some cases of physical interest it is essential to study finite systems, whose properties near a phase transition are characterized by a finite-size scaling (FSS) behavior \cite{fisher_barber,cardy1,privman,PELISSETTO_VICARI,campostrini_pelissetto_vicari}. The understanding of finite-size effects at phase transitions has great phenomenological importance, since it is crucial to interpret experiments as well as numerical investigations of finite systems at the transition point. 
In the case of a continuous phase transition, FSS is characterized by power laws with universal critical exponents, in the sense that they depend only on global features of the system (e.g. symmetries of the hamiltonian, number of spatial dimensions) and are independent of the geometry and of the boundary conditions. 

For a first-order transition (FOT) the FSS behavior may instead depend on the boundary conditions and on the geometry \cite{privman_fisher,fisher_privman_oNsymmetrybreaking}: in the case of periodic boundary conditions (PBC), which is the most studied in literature \cite{nienhuis_nauenberg,challa_landau_binder,binder}, the finite-size effects are characterized by a power-law behavior, whose exponents are related to the space dimension of the system. These effects are different if general boundary conditions are considered, as noted in Refs. \cite{Panagopoulos_pelissetto_vicari_anomalousfss,panagopoulos_vicari}; moreover, recent studies of quantum FOTs have also reported a significant dependence on the boundary conditions \cite{pelissetto_rossini_vicari}. It is then interesting to show how the scaling behavior of a finite system is modified in correspondence of different boundary conditions, both for equilibrium and off-equilibrium properties.\\

In this paper, we study the role of the boundary conditions in the context of the simplest classical system exhibiting a magnetic FOT, i.e. the two-dimensional (2D) Ising model in the low-temperature phase with an applied external magnetic field. In particular, we analyze both static and dynamic behaviors at the transition in the cases of opposite fixed boundary conditions (OFBC) and open boundary conditions (OBC), when a purely relaxational dynamics is considered. In the OFBC case, the boundaries are chosen in order to favor the formation of a planar interface separating the phases. This interface moves within the lattice, and gives rise to a dynamics that is different compared to the PBC case. Indeed, this is related to the dependence of the equilibrium relaxational dynamics at FOTs on the boundary conditions. In the PBC case, systems of size $L$ are characterized by an exponential dynamics due to an exponentially large tunneling time $\tau(L)\sim e^{\sigma L}$ between the coexisting phases. The dynamics of other types of boundary conditions favoring the formation of an interface may lead to a power-law behavior, as in the case of OFBC. Instead, in the OBC case the boundaries do not favor any specific phase within the system. Here we show that an appropriate dynamic finite-size scaling (DFSS) theory can be developed, provided that the system is in the coexistence region; in particular, the time scale of the dynamics $\tau(L)$ is exponential in $L$, as it happens for PBC, but with a different coefficient in the exponent, that can be justified by looking at the typical configurations promoted by OBC in the coexistence region.

The paper is organized as follows. In Sec. \ref{2dising_section2} we consider the 2D Ising model, defining the relevant observables and specifying the details of the relaxational dynamics. The equilibrium properties in the OFBC case are analyzed in Sec. \ref{equilibrium_OFBC}, where we recall briefly the known features of FSS for neutral boundary conditions. The dynamic scaling behavior for both types of boundary conditions is analyzed, and compared with the PBC case, in Sec. \ref{dynamic_OBC}. In Sec. \ref{conclusions} we summarize and present our conclusions. Appendix \ref{binning method} contains some details of our numerical estimates of the autocorrelation time.

\section{\label{2dising_section2}The two-dimensional Ising model}
We consider the 2D Ising model on a square lattice $L\times L$ in the presence of an external magnetic field $h$. The hamiltonian is
\begin{equation}
H=-\sum_{\langle i,j\rangle}s_is_j-h\sum_is_i,
\end{equation}
where $s_i\in\{-1,1\}$ and the symbol $\langle i,j\rangle$ denotes a nearest neighbor pair of spins. The model undergoes a paramagnetic-ferromagnetic transition for $h=0$ and $T=T_c$, with \cite{huang}
\begin{equation}
T_c=\frac{2}{\ln{(1+\sqrt{2})}}, \;\;\;\;\; \beta_c\equiv\frac{1}{T_c}.
\end{equation}
For $h\rightarrow 0$, $T<T_c$ the system is spontaneously magnetized in the thermodynamic limit, and the expression for the spontaneous magnetization is \cite{yang}
\begin{equation}
m_0(\beta)=[1-\sinh{(2\beta)}^{-4}]^{1/8}.
\label{spontaneous_magnetization_ising}
\end{equation}
In the following we also need the interface tension $\kappa$ at fixed $\beta$, known exactly for this model \cite{zia_avron}:
\begin{equation}
\kappa(\beta)=2+\frac{\ln{[\tanh{(\beta)}}]}{\beta}.
\end{equation}
Here we are going to investigate the finite-size behavior of the average magnetization density, defined as
\begin{equation}
M=\frac{1}{L^2}\sum_is_i.
\end{equation}
We define also the corresponding renormalized magnetization $m_r$ as
\begin{equation}
m_r=\frac{M}{m_0},
\end{equation}
and also its average over different dynamic histories
\begin{equation}
m_r(t,r_1,L)=\frac{\langle M(t)\rangle}{m_0}.
\end{equation}
In our analysis a purely relaxational dynamics at fixed $T<T_c$ and fixed $h$ is considered: the spins interact with the external field in such a way that the total spin is not conserved and the system relaxes to its thermodynamic equilibrium state.\\
Since we will also analyze the dynamical behavior of $M$ with respect to the time $t$, given a real number $\mu\in(-1,1)$, we also define the first-passage time (FPT) $t_f(\mu)$ as the smallest time such that
\begin{equation}
M[t_f(\mu)]\equiv\mu m_0,
\end{equation}
then we can consider its average
\begin{equation}
T_f(\mu,r_1,L)=\langle t_f(\mu)\rangle.
\end{equation}
This last quantity will be relevant only in the DFSS analysis, i.e. in the OBC case, and will play no role in the OFBC case.

\section{\label{equilibrium_OFBC}Equilibrium behavior in the coexistence region}

In a finite square box of linear size $L$, the behavior of the system depends on the boundary conditions. The equilibrium FSS behavior for neutral boundary conditions, i.e. boundaries that preserve the $\mathbb{Z}_2$ inversion symmetry of the model, like the PBC, has been already studied in earlier works \cite{privman}. In order to better appreciate the new features of FSS for the OFBC case, it is instructive to first briefly summarize the known features of FSS for neutral boundaries, and this will be done in the next two subsections. Then the results of our studies for the OFBC case are reported and compared with the known theory.
\subsection{\label{knownFSS_PBC and OBC} Finite-size scaling for neutral boundary conditions}

In a finite system of size $L$ in the low-temperature phase the variation of the magnetization $m$ with the external field $h$ is perfectly smooth: rather than an infinitely steep variation with $h$, $m$ has a large but finite slope. In the case of neutral boundary conditions preserving the $\mathbb{Z}_2$ inversion symmetry, like PBC and OBC, a FSS theory can be developed in the case of a field-driven FOT for a ferromagnetic system below the critical temperature \cite{Binder1981,binder}. 

For $h\rightarrow 0^\pm$ and $L\gg\xi$, where $\xi$ is the correlation length of the low-temperature phase, the probability distribution of the magnetization $P_L(m)$ is a sum of two gaussians, centered at $m_0$ and $-m_0$. For $h\neq0$ the probability distribution is again a sum of two gaussians, but now centered around the shifted values $\pm m_0+h\chi_m$, where $\chi_m$ is the magnetic susceptibility: in this case the weights of the two peaks are different, i.e. one of the phases is favored by $h$ according to its sign.

From $P_L(m)$ we can obtain the average magnetization
\begin{equation}
\langle m\rangle=h\chi_m+m_0\tanh{\beta h m_0L^d}.
\end{equation}
For small $h$, the relevant scaling variable in the 2D static case is 
\begin{equation}
r_1=hL^2,
\end{equation}
so the appropriate universal behavior is observed when $h\rightarrow 0$, $L\rightarrow\infty$ at fixed $r_1$. In particular, when the FSS limit is considered the magnetization per site $m$ becomes
\begin{equation}
m=m_0\;f_m(r_1), \;\;\;\;\; f_m(r_1)=\tanh{(\beta m_0r_1)}.
\label{equilibrium value magnetization}
\end{equation}
If the value of $r_1$ is finite then $m\neq |m_0|$, indicating that both free-energy minima contribute to equilibrium properties, a sign of the fact that the system is always in the coexistence region.

\subsection{\label{fseffects_coexregion} Surface effects in the coexistence region}
The double-gaussian approximation is reasonable only near $m=\pm m_0$, while it gives an underestimate of the real value of $P_L(m)$ in the interval $-m_0\ll m \ll m_0$. In this region the probability distribution is dominated by configuration corresponding to the two-phase coexistence in the system. The correct value of $P_L(m)$ for $m\approx 0$ can be obtained by considering interface contributions in the free energy $\mathcal{F}_L(m)$ and then writing
\begin{equation}
P_L(m)=\frac{e^{-\beta\mathcal{F}_L(m)}}{Z_L},
\end{equation}
where $Z_L$ is the partition function. Let us denote with 
\begin{equation}
\Delta\mathcal{F}_L\equiv\mathcal{F}_L(m\approx0)-\mathcal{F}_L(m\approx m_0)
\end{equation}
the free energy difference between the configuration characterized by the phase coexistence, i.e. $m\approx0$, and the configuration characterized by the presence of a single phase, i.e. $m\approx m_0$. Then the interface contributions are \cite{Binder1981}
\begin{equation}
\Delta\mathcal{F}_L=2L^{d-1}\kappa, \;\;\;\;\; \text{for PBC},
\end{equation}
\begin{equation}
\Delta\mathcal{F}_L=L^{d-1}\kappa, \;\;\;\;\; \text{for OBC},
\end{equation}
in the $d$-dimensional case. These differences are computed taking into account the typical configurations of the system in the coexistence region, as we shall see in the next sections.

It is worth mentioning that in these specific cases there is no shift of the transition, due to inversion symmetry. If boundary conditions that explicitly break this symmetry are considered, the transition point is shifted to a non-trivial value $h(L)\propto L^{-1}$ \cite{Privman1990}.

\subsection{\label{subsection_scaling_OFBC}Magnetization profile for opposite boundaries}
We study the equilibrium scaling behavior of finite systems at magnetic FOTs, when the boundary conditions at two opposite sides of the system generate an interface. For the 2D Ising model, these boundary conditions are realised by considering OFBC along the $x$-direction, i.e. a column of positive spins (on the right) favors the phase with positive magnetization while a column of negative spins (on the left) favors the other one, and PBC in the $y$-direction.

We analyze the equilibrium scaling behavior of the average magnetization $\langle M\rangle$ in terms of the size $L$. As for PBC, these boundary conditions preserve the inversion symmetry $h\rightarrow-h$ and then we expect that the correct scaling variable for small $h$ is $r_1=hL^2$ in the 2D case. 

We perform various Monte Carlo (MC) simulations with different fixed value of $r_1$ and fixed $T=0.8T_c$, using a standard Metropolis single-spin flip algorithm. If the scaling variable is correct we should observe that
\begin{equation}
m_r=g_m(r_1),
\label{equation_scaling_OFBC}
\end{equation}
where $g_m$ is a scaling function.

\begin{figure}[h]
	\centering
	\includegraphics[width=0.8\linewidth]{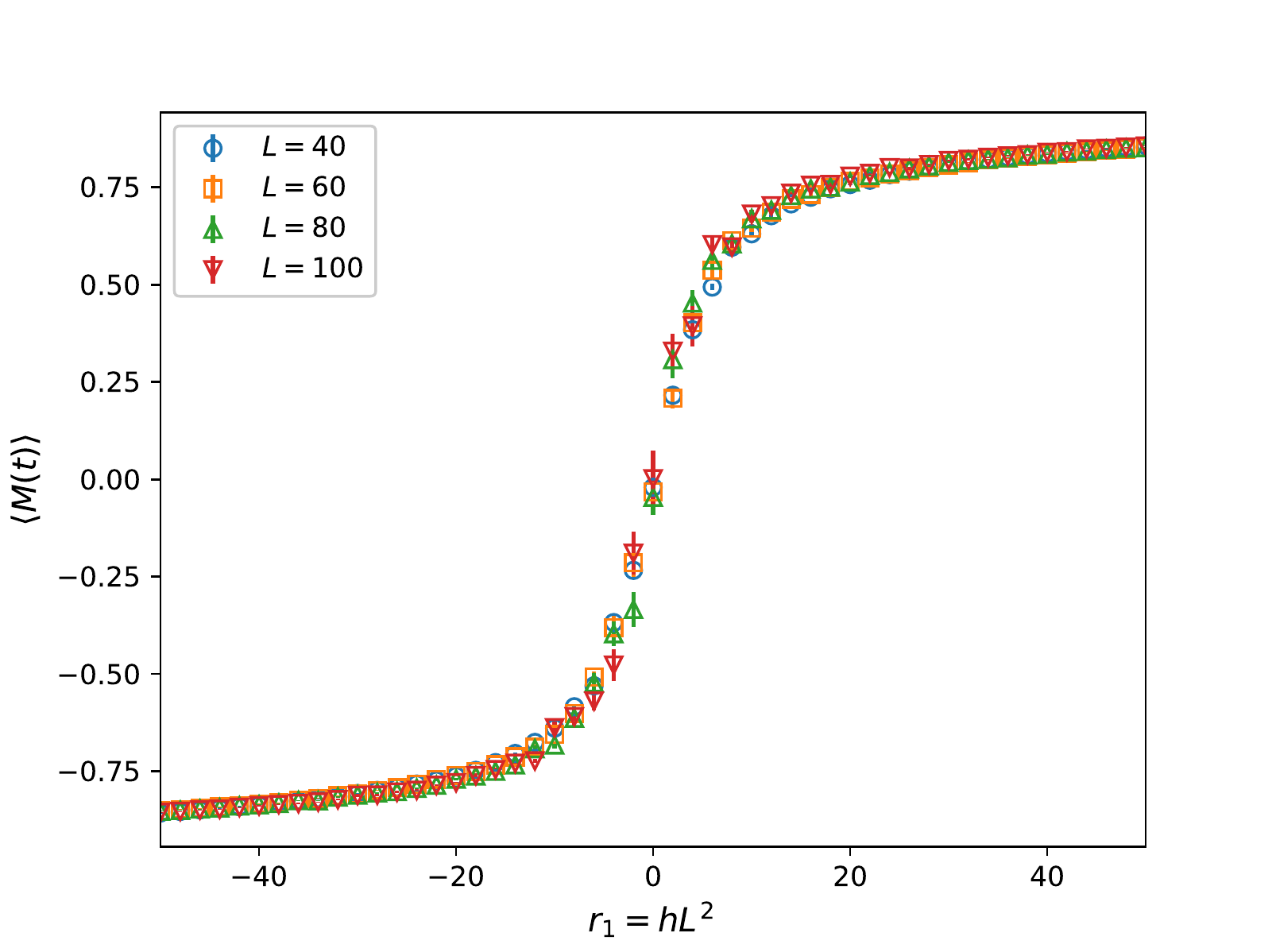}
	\caption{Scaling plot of the average renormalized magnetization $m_r$ versus the scaling variable $r_1$ for different sizes $L$.}
	\label{scaling_OFBC}
\end{figure}

\begin{figure}[h]
	\centering
	\includegraphics[width=0.7\linewidth]{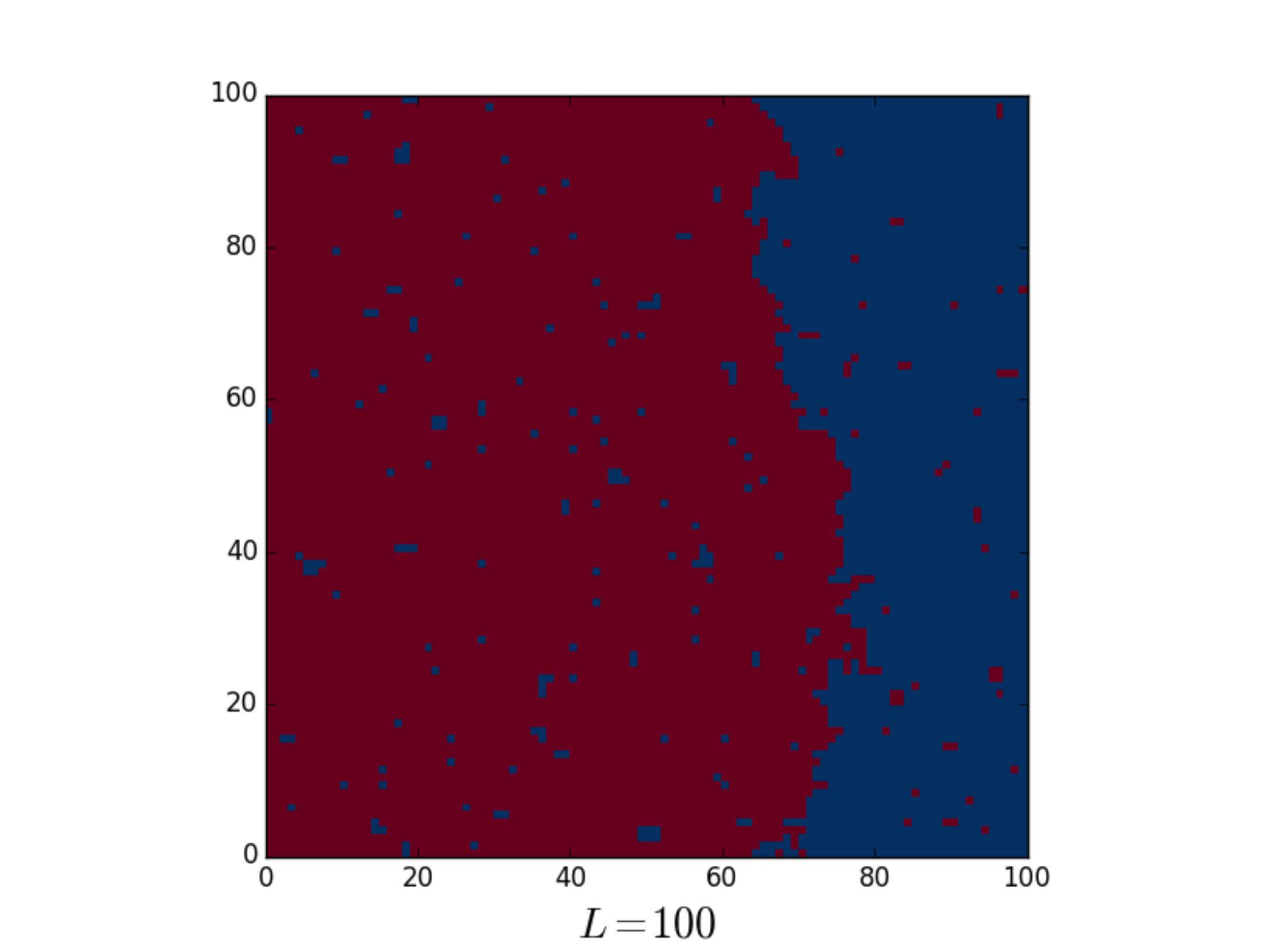}
	\caption{Typical configuration with $L=100$ and $m\approx-0.4$. Red regions are characterized by a negative value of the magnetization, blue regions by a positive value.}
	\label{config_OFBC}
\end{figure}

Data are shown in fig. \ref{scaling_OFBC} and we clearly see that the scaling is optimal. We can interpret this in terms of the interface generated by the boundary conditions: in the coexistence region the typical configuration has the phase with positive magnetization on the right side of the lattice, and negative magnetization on the left side, as in fig. \ref{config_OFBC}.
\\
Let us denote with $X_i$ the location of the interface along the $x$-axis and with $A_\pm$ the areas of positive/negative magnetization regions. Setting $x=0$ in the center of the lattice, we have that $X_i\in[-L/2,L/2]$. Furthermore
\begin{equation}
m\left(X_i=\mp\frac{L}{2}\right)=\pm m_0.
\end{equation}
We are able to write the magnetization as a function of $A_\pm$, indeed
\begin{equation}
m=m_0(2a_+-1), \;\;\;\;\; a_+\equiv\frac{A_+}{L^2}.
\end{equation}
We can relate $m(a_+)$ to the coordinate $X_i$: if we suppose $X_i>0$, i.e. that the interface is located in the right half of the lattice, then
\begin{equation}
a_+=\frac{1}{L}\left(\frac{L}{2}-X_i\right),
\end{equation}
therefore
\begin{equation}
m=m_0\left[\frac{2}{L}\left(\frac{L}{2}-X_i\right)-1\right]=-\frac{2m_0}{L}X_i.
\end{equation}
Thus, considering also eq.(\ref{equation_scaling_OFBC}), it follows that $X_i=\widetilde{g}_m(r_1)$, where $\widetilde{g}_m$ is another scaling function such that $\widetilde{g}_m\propto g_m$.

The connection between $r_1$ and $X_i$ is represented by this proportionality: if we move towards positive values of $r_1$ the interface is moved to the left; conversely, if we move towards negative values of $r_1$ the interface will be moved to the right. In this sense the interface follows the scaling variable $r_1$.

\section{\label{dynamic_OBC}Dynamic scaling behavior in the coexistence region}
After having analyzed the equilibrium FSS properties in the OFBC case, we investigate the dynamical properties of the system in the coexistence region both for OBC and OFBC. In particular, we analyze the dynamic scaling behavior of the 2D Ising model in the OBC case and we give an estimate of the equilibrium dynamic exponent $z$ in the OFBC case. Before showing our numerical results, we briefly recall what happens in the PBC case and give some physical arguments for the scaling behavior of the dynamic time scale. Moreover, we also provide some details about the DFSS theory used to the describe the system in the coexistence region.

\subsection{\label{arguments_PBCT}Time scale for periodic boundary conditions}
The coexistence region for the 2D Ising model is represented by the line $h=0$, $T<T_c$ in the relative phase diagram. Close to this segment, physical observables show a scaling behavior in terms of $h$ and $L$ that depends on the boundary conditions. To extend FSS to the dynamic case it is necessary to identify the time scale of the dynamics. Recalling that in the static case $h\propto L^{-2}$ at fixed $r_1$, we note that $h\rightarrow0$ in the scaling limit $L\rightarrow\infty$. This means that the system is in the coexistence region in this limit, thus as a relevant time scale we can consider the one that controls the large-time dynamic behavior for $h=0$. 

If the considered boundary conditions are symmetric, for $T<T_c$ the largest relaxation times are associated with flips of the magnetization. In the PBC case the typical configuration is characterized by the presence of two interfaces separating the coexistent phases, while configurations with spherical droplets are unstable in this region \cite{berg_hansmann_neuhaus, bray}. Since the time needed to observe a complete reversal of the magnetization is proportional to $e^{2\beta\kappa L}$ \cite{miyashita_takano,binder_heermann}, the correct time scale is 
\begin{equation}
\tau(L)=L^\alpha e^{2\kappa\beta L},
\end{equation}
where $\alpha$ is an appropriate exponent  and the factor of two in the exponent is due to the presence of two interfaces separating the phases.

\subsection{\label{arguments_OBC} Time scale for open boundary conditions}

The arguments used in the PBC case can be applied also for OBC; however these boundary conditions give rise to different configurations in the lattice. In particular, the typical configurations in the coexistence region are characterized by the presence of spherical domains or by the presence of a single planar interface. These two types of configurations are characterized by different values of the magnetization, and there is a critical value that separates them. In the 2D case, this value is \cite{Binder1981}
\begin{equation}
m_c=\left(1-\frac{1}{2\pi}\right)M_0\approx0.841M_0,
\end{equation}
where $M_0=\pm1$ is the value of the magnetization in one of the pure phases of the system. We clearly see that, since $M_0\approx m_0$, in the region $\Lambda=[-0.841m_0,0.841m_0]$ configurations characterized by the presence of a single interface should be energetically preferred.

Assuming that the relevant mechanism for the generation of configurations with two coexisting phases is the creation of domain walls parallel to the lattice axis, we therefore expect that the correct time scale for OBC is
\begin{equation}
\tau(L)=L^\alpha e^{\kappa\beta L}
\label{timescaleOBC}
\end{equation}
and we define
\begin{equation}
r_2=\frac{t}{\tau(L)}
\end{equation}
as our dynamic scaling variable. 

The behavior in this case is expected to be ruled by the same mechanism of the PBC case, but with different time scales since different configurations are promoted. For this reason, in the next section we briefly summarize what is know for PBC, in order to extend the DFSS also to the OBC case.

\subsection{\label{cg_OBC}Coarse-grained flip dynamics for PBC}
In the PBC case, physical observables such as $m_r(t,r_1,L)$ and $T_f(\mu,r_1,L)$ are expected to show a scaling behavior in terms of the scaling variables we have identified. 
In particular we expect
\begin{equation}
T_f(\mu,r_1,L)\approx\tau(L)f_T(r_1,\mu),
\end{equation}
\begin{equation}
m_r(t,r_1,L)\approx g_m(r_1,r_2).
\end{equation}
These scaling function can be exactly predicted by modeling the dynamics of the system as a simple two-level dynamics, provided that time scales of the order of $\tau(L)$ are considered. Indeed, in this case what we observe for a single dynamic history is that the system oscillates between the two Ising phases, characterized by the values of the spontaneous magnetization equal to $\pm m_0$. The fact that the time scales are of the order of $\tau(L)$ allows us to consider the flip of the magnetization essentially instantaneous.

The dynamics, assuming it is Markovian, is completely parametrized by the rates
\begin{equation}
P[M(t)=\mp m_0\rightarrow M(t+dt)=\pm m_0]=I_{\pm}dt,
\end{equation}
being $P(\cdot)$ the probability of such a transition from a phase to the other one.\\
Considering different dynamic realizations of the process, it can also be shown that the average renormalized magnetization is given by
\begin{equation}
m_r(t)=\frac{I_+-I_-}{I_++I_-}-\frac{2I_+}{I_++I_-}e^{-(I_++I_-)t},
\end{equation}
being the rates related by
\begin{equation}
\frac{I_-}{I_+}=e^{-2\beta m_0r_1}.
\end{equation}
Since in the considered limit the flips are instantaneous, $T_f(\mu,r_1,L)$ is expected to be independent of $\mu$, i.e. $T_f(\mu,r_1,L)\approx\tau_f(r_1,L)$. Moreover, the rate $I_+$ and the FPT are related via
\begin{equation}
I_+^{-1}=\tau_f(r_1,L).
\end{equation}
This allows us to write, in the scaling limit
\begin{equation}
m_r(t)\approx f_m(r_1)-[1+f_m(r_1)]e^{-t/T_i},
\label{renormalized_magnetization_OBC}
\end{equation}
\begin{equation}
T_f(\mu,r_1,L)\approx\tau(L)g_\tau(r_1),
\end{equation}
where
\begin{equation}
T_i\equiv\frac{\tau_f(r_1,L)}{1+e^{-2\beta m_0r_1}}
\end{equation}
and $g_\tau(r_1)$ is a scaling function satisfying
\begin{equation}
\frac{g_\tau(r_1)}{g_\tau(-r_1)}=e^{-2\beta r_1m_0}.
\end{equation}
These equations tell us two things: the first is that the time scale $\tau(L)$ can be evaluated once the mean FPT is known at fixed $\mu$ and $r_1$, while the second is that the behavior of the average renormalized magnetization is exponential in a proper variable, which is nothing but the time $t$ rescaled with the mean FPT.

This theory was succesfully analyzed numerically in Ref. \cite{pelissetto_vicari_DFSS}; however, we expect that the DFSS applies also in the OBC case, since these boundaries do not favor any specific phase of the system. In particular, we expect that the dynamics is again exponential, with a time scale that is a half of the one characterizing the PBC case. Moreover, we argue that the behavior of the average renormalized magnetization is again exponential, but in principle with a different scaling function.

\subsection{\label{results_OBC}Open boundaries: data and results in the coexistence region}
To see if our predictions are correct, we perform MC simulations for different values of $L\in[20,40]$ and $r_1$, using again the Metropolis algorithm to implement a purely relaxational dynamics at fixed $T=0.9T_c$. 

The mean FPT is independent of $\mu$ for $L\rightarrow\infty$, thus we can fix $\mu=0.9$ as our reference value.
\begin{figure}[h]
	\centering
	\includegraphics[width=0.7\linewidth]{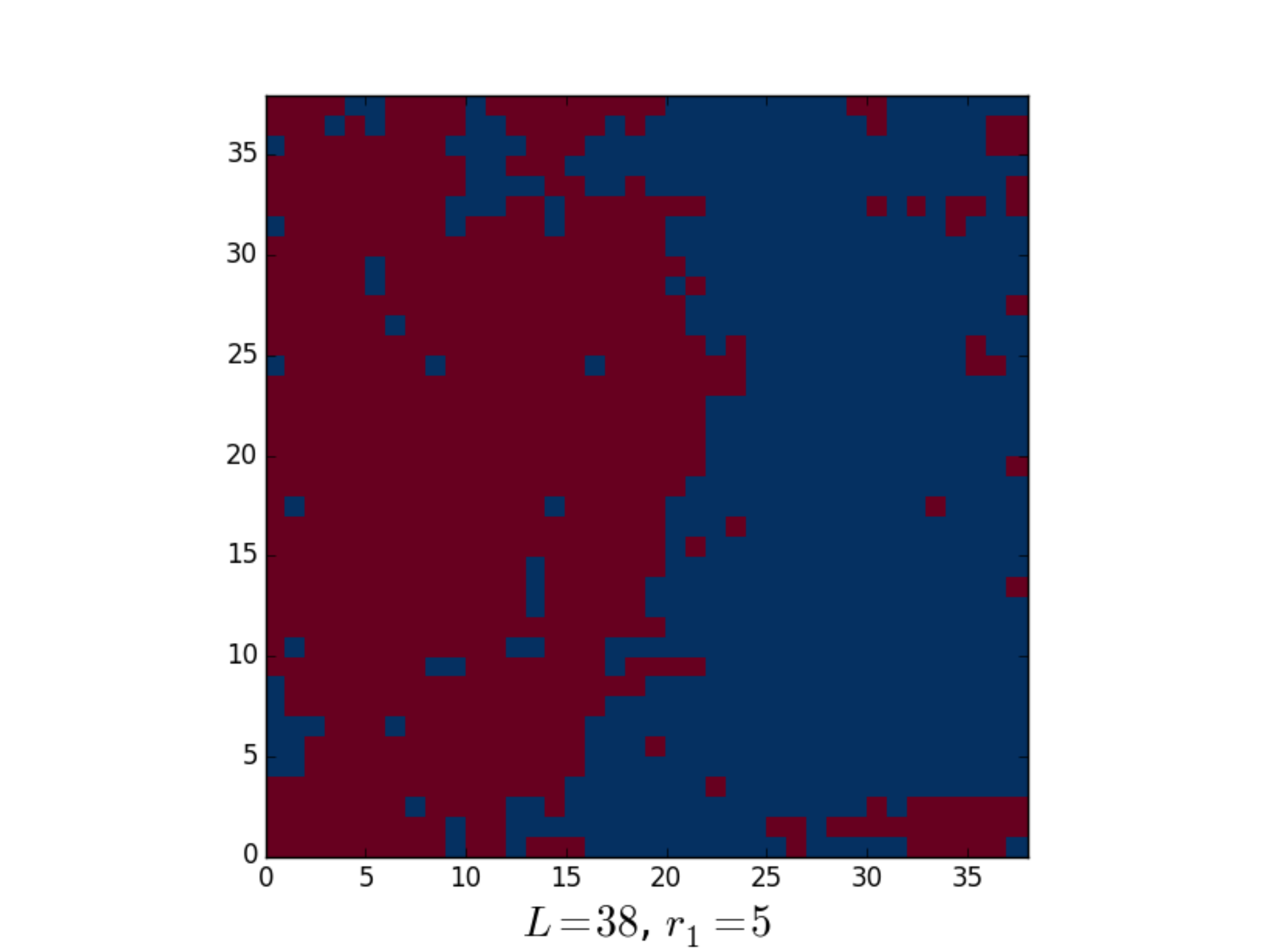}
	\caption{Typical configuration with $L=38$ and $r_1=5$ in the coexistence region. Red regions are characterized by a positive value of the magnetization, blue regions by a negative value.}
	\label{data5r1_75r1}
\end{figure}
We first observe what is the typical configuration in the coexistence region, i.e. when $\langle M(t)\rangle\approx0$. An example is reported in fig. \ref{data5r1_75r1}: we can see that opposite phases are separated by a single (approximately) planar interface, without spherical domains. Therefore on the basis of what obtained with MC simulations we can neglect spherical droplets and consider eq. (\ref{timescaleOBC}) as the time scale for OBC.

Moreover, by looking at the MC histories, we observe that $M(t)$ oscillates between two phases that are characterized by a value of the magnetization which is slightly different from the one given by eq. \eqref{spontaneous_magnetization_ising} at the given $\beta=1/T$. This value $\tilde{m}_0$ turns out to be smaller than $m_0$, in particular $\tilde{m}_0\approx0.81$ while $m_0\approx0.895$ by using eq. \eqref{spontaneous_magnetization_ising}. This is due to finite-size effects, since we are using values of $L$ up to $L=40$ in our simulations, while eq. \eqref{spontaneous_magnetization_ising} holds in the thermodynamic limit. Moreover, the fact that now OBC are considered renders more pronounced this discrepancy: the PBC case \cite{pelissetto_vicari_DFSS} is obviously different, since they minimize the effects due to the finiteness of the lattice.

We want to check the size dependence of this time scale. Numerically we compute $T_f(\mu=0.9, r_1, L)$ for different values of $r_1$, precisely $r_1=5$, $r_1=7.5$ and $r_1=10$. Initially we fit these data to the unbiased ansatz
\begin{equation}
\log{T_f(\mu=0.9, r_1, L)}=aL+c,
\end{equation}
i.e. not taking into account the power-law corrections. If eq. (\ref{timescaleOBC}) holds we should find a value  $a\approx\kappa\beta$, which for the chosen value of $T$ is $\kappa\beta\approx0.189514$.
\begin{figure}[h]
	\centering
	\includegraphics[width=0.8\linewidth]{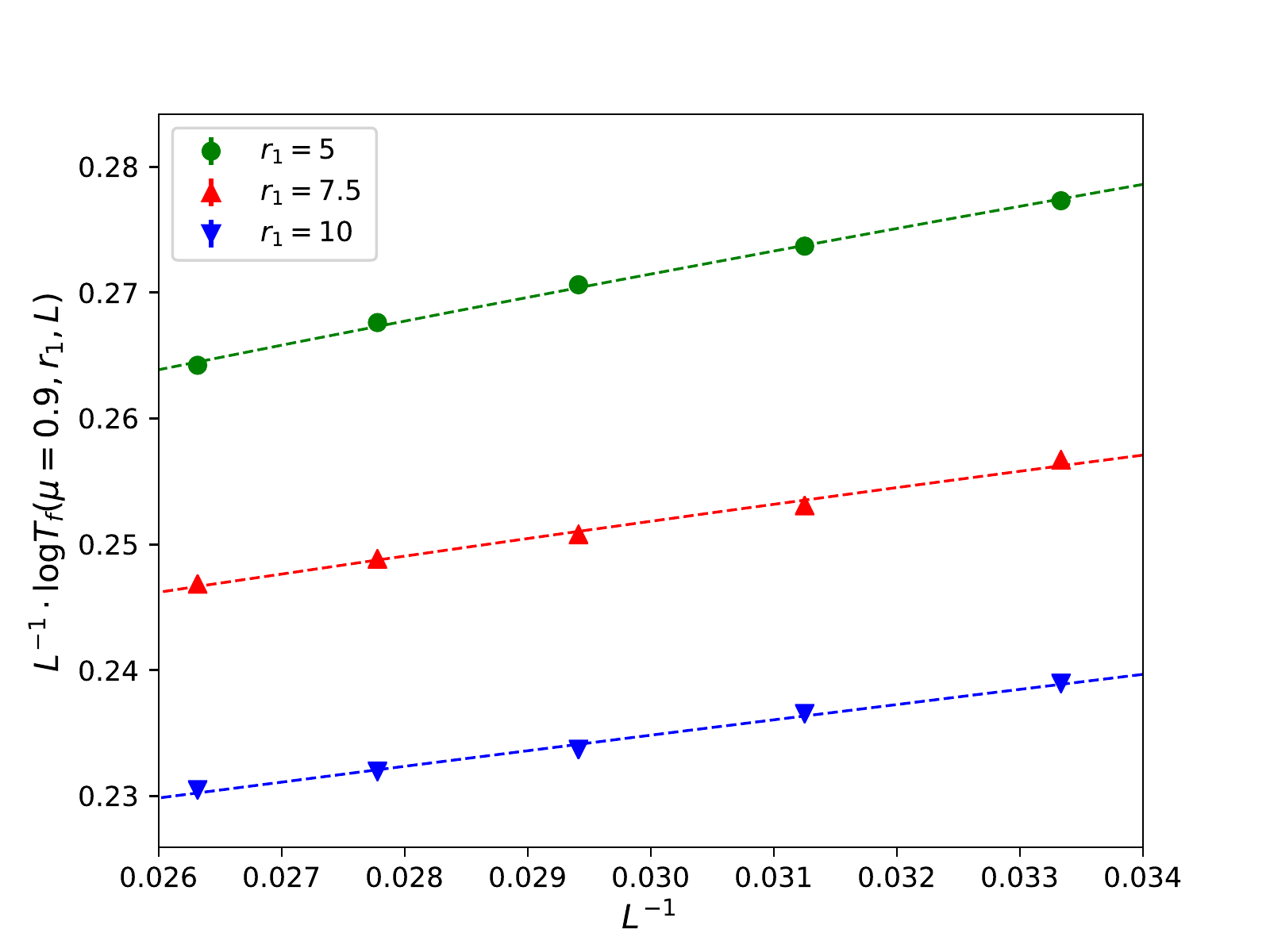}
	\caption{Data and fits (dashed lines) of mean FPTs for different $r_1$ as functions of $L^{-1}$. These fits are those with $a=\sigma$. The error bars are reported, although they are small compared to the size of the markers.}
	\label{OBC_dataplot}
\end{figure}
The results of the fits are reported in Table \ref{fit_parameters_freesigma}.
\begin{table}[h]
	\setlength{\tabcolsep}{9pt}
	\begin{center}
		\begin{tabular}{ccccc}
			\hline 
			\hline
			$r_1$ & Range of $L$ & $a$ & $c$ & $\chi^2/\mathrm{ndof}$ \\  
			\hline
			5 & $[24,38]$ & $0.218(1)$ & $1.76(3)$ & $14.5/6$\\
			& $[26,38]$ & $0.218(1)$ & $1.76(5)$ & $14.0/5$\\
			& $[28,38]$ & $0.218(2)$ & $1.75(7)$ & $14.3/4$\\
			& $[30,38]$ & $0.216(2)$ & $1.85(7)$ & $5.8/3$\\
			\hline
			7.5 & $[24,38]$ & $0.207(1)$ & $1.49(4)$ & $18.0/6$\\
			& $[26,38]$ & $0.208(1)$ & $1.45(4)$ & $12.7/5$\\
			& $[28,38]$ & $0.210(1)$ & $1.39(4)$ & $4.9/4$\\
			& $[30,38]$ & $0.211(2)$ & $1.36(6)$ & $4.4/3$\\
			\hline
			10 & $[24,38]$ & $0.192(1)$ & $1.42(4)$ & $29.1/6$\\
			& $[26,38]$ & $0.193(1)$ & $1.36(5)$ & $16.6/5$\\
			& $[28,38]$ & $0.195(2)$ & $1.31(5)$ & $4.9/4$\\
			& $[30,38]$ & $0.197(2)$ & $1.23(5)$ & $3.6/3$\\
			\hline
			\hline
		\end{tabular}
	\end{center}
	\caption{Fit parameters for different ranges of $L$ and different $r_1$; ndof is the number of the degrees of freedom of the fit.}
	\label{fit_parameters_freesigma}
\end{table}
\\ Concerning $a$, we observe a trend with the size $L$ of the system: when lower values of $L$ are discarded $a$ grows, except for the case $r_1=5$. Furthermore, it seems that there is a dependence also on the value of $r_1$: as it increases $a$ becomes smaller. All these values of $a$ are not distant from the theoretical value $\beta\kappa$, but they are not compatible with each other within the errors. Despite this, we can conclude that the analysis is consistent with an exponentially slow dynamics, with an exponent close to the expected value.

To check if there are significant power-law corrections to this exponential behavior we can fix $a=\beta\kappa$ in eq. (\ref{timescaleOBC}) and try to make a fit of the form
\begin{equation}
\log{[T_f(\mu=0.9,r_1,L)e^{-\beta\kappa L}]}=\alpha\log{L}+c.
\label{corr_timescaleOBC}
\end{equation}
Parameters for different values of $r_1$ are reported in Table \ref{fit_parameters_fixedsigma}. We observe again a trend with the size $L$: as the latter increases also $\alpha$ increases. It seems that for $\alpha$ these is a dependence on the value of $r_1$, as happens for the estimated exponent $a$. Moreover, big values of $\chi^2/\mathrm{ndof}$ suggest that there are significant corrections to scaling: therefore, it is difficult to give a final reliable estimate of $\alpha$. Despite that, if we look at fig. \ref{OBC_dataplot} is clear that the dynamics is consistent with a mainly exponential behavior with power-law corrections, even if here additional corrections to scaling are not considered.
\begin{table}[h]
	\setlength{\tabcolsep}{9pt}
	\begin{center}
		\begin{tabular}{ccccc}
			\hline 
			\hline
			$r_1$ & Range of $L$ & $\alpha$ & $c$ & $\chi^2/\mathrm{ndof}$ \\ 
			\hline
			5 & $[24,38]$ & $0.88(3)$ & $-0.4(1)$ & $13.7/6$\\
			& $[26,38]$ & $0.92(4)$ & $-0.5(1)$ & $10.6/5$\\
			& $[28,38]$ & $0.96(5)$ & $-0.6(2)$ & $7.2/4$\\
			& $[30,38]$ & $0.89(5)$ & $-0.4(2)$ & $3.5/3$\\
			\hline
			7.5 & $[24,38]$ & $0.53(5)$ & $0.2(2)$ & $33.3/6$\\
			& $[26,38]$ & $0.58(5)$ & $0.03(20)$ & $21.5/5$\\
			& $[28,38]$ & $0.67(5)$ & $-0.3(2)$ & $8.1/4$\\
			& $[30,38]$ & $0.72(7)$ & $-0.4(3)$ & $6.3/3$\\
			\hline
			10 & $[24,38]$ & $0.07(5)$ & $1.2(1)$ & $31.9/6$\\
			& $[26,38]$ & $0.13(5)$ & $1.0(2)$ & $18.9/5$\\
			& $[28,38]$ & $0.19(6)$ & $0.8(2)$ & $11.6/4$\\
			& $[30,38]$ & $0.28(6)$ & $0.5(2)$ & $4.2/3$\\
			\hline
			\hline
		\end{tabular}
	\end{center}
	\caption{Parameters for the fits at fixed $a=\beta\kappa$ for different ranges of $L$ and different $r_1$; ndof is the number of the degrees of freedom of the fit.}
	\label{fit_parameters_fixedsigma}
\end{table}

\begin{figure}[h]
	\centering
	\includegraphics[width=0.8\linewidth]{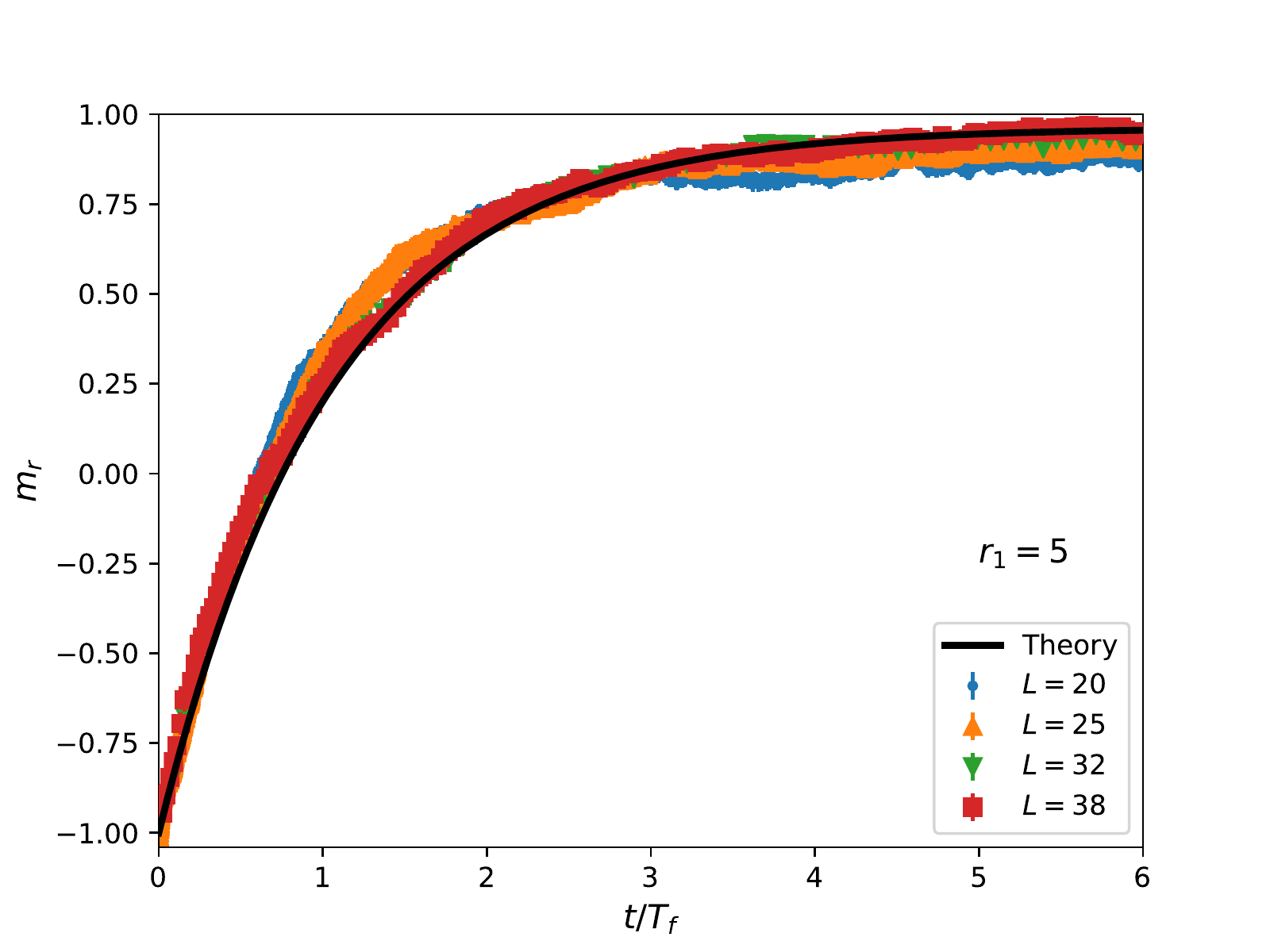}
	\caption{Renormalized magnetization $m_r=M/\tilde{m}_0$ as a function of $t/T_f(0.9,r_1,L)$ for $r_1=5$.}
	\label{OBC_mr5}
\end{figure}

\begin{figure}[h]
	\centering
	\includegraphics[width=0.8\linewidth]{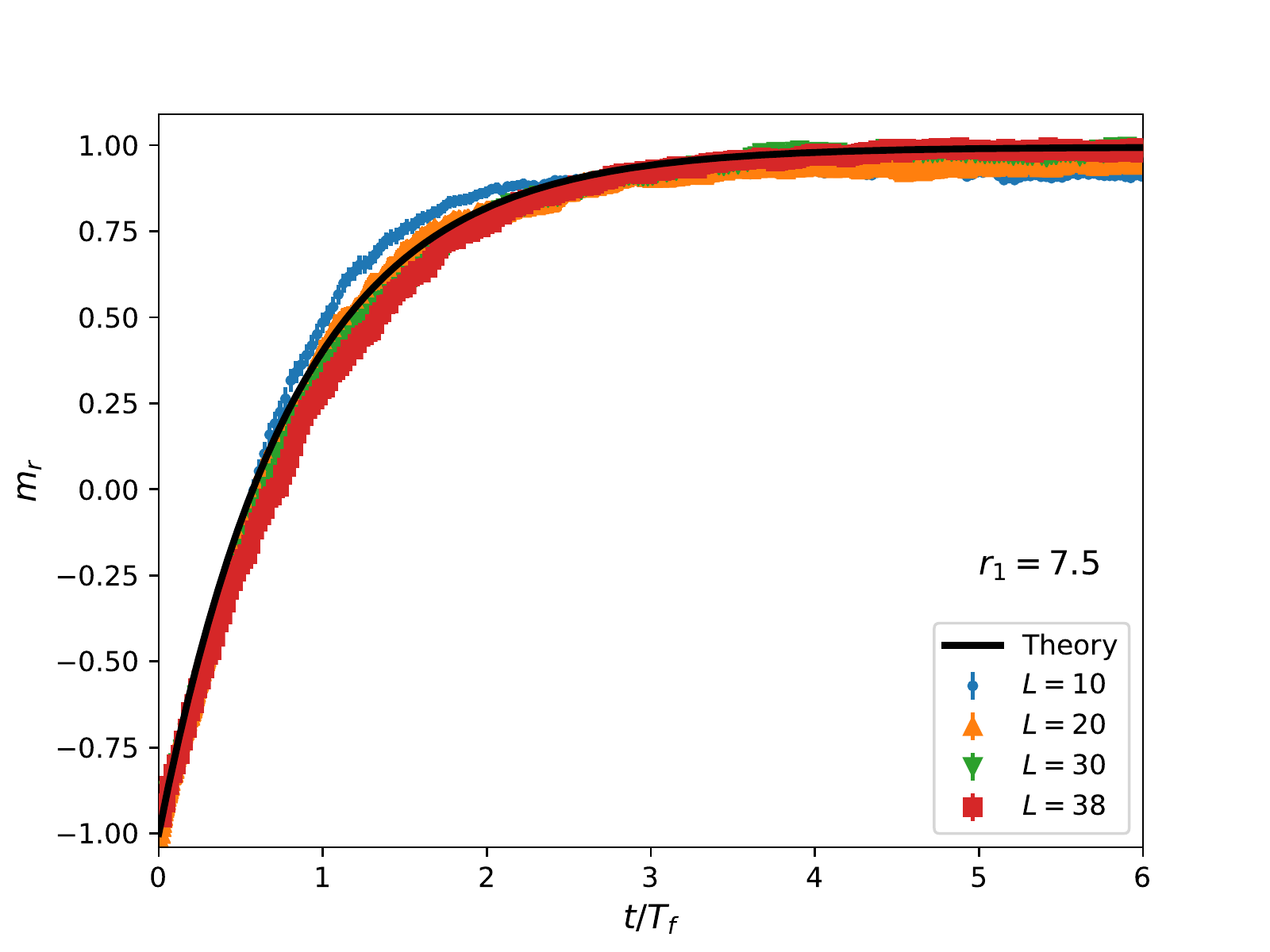}
	\caption{Renormalized magnetization $m_r=M/\tilde{m}_0$ as a function of $t/T_f(0.9,r_1,L)$ for $r_1=7.5$.}
	\label{OBC_mr75}
\end{figure}

\begin{figure}[h]
	\centering
	\includegraphics[width=0.8\linewidth]{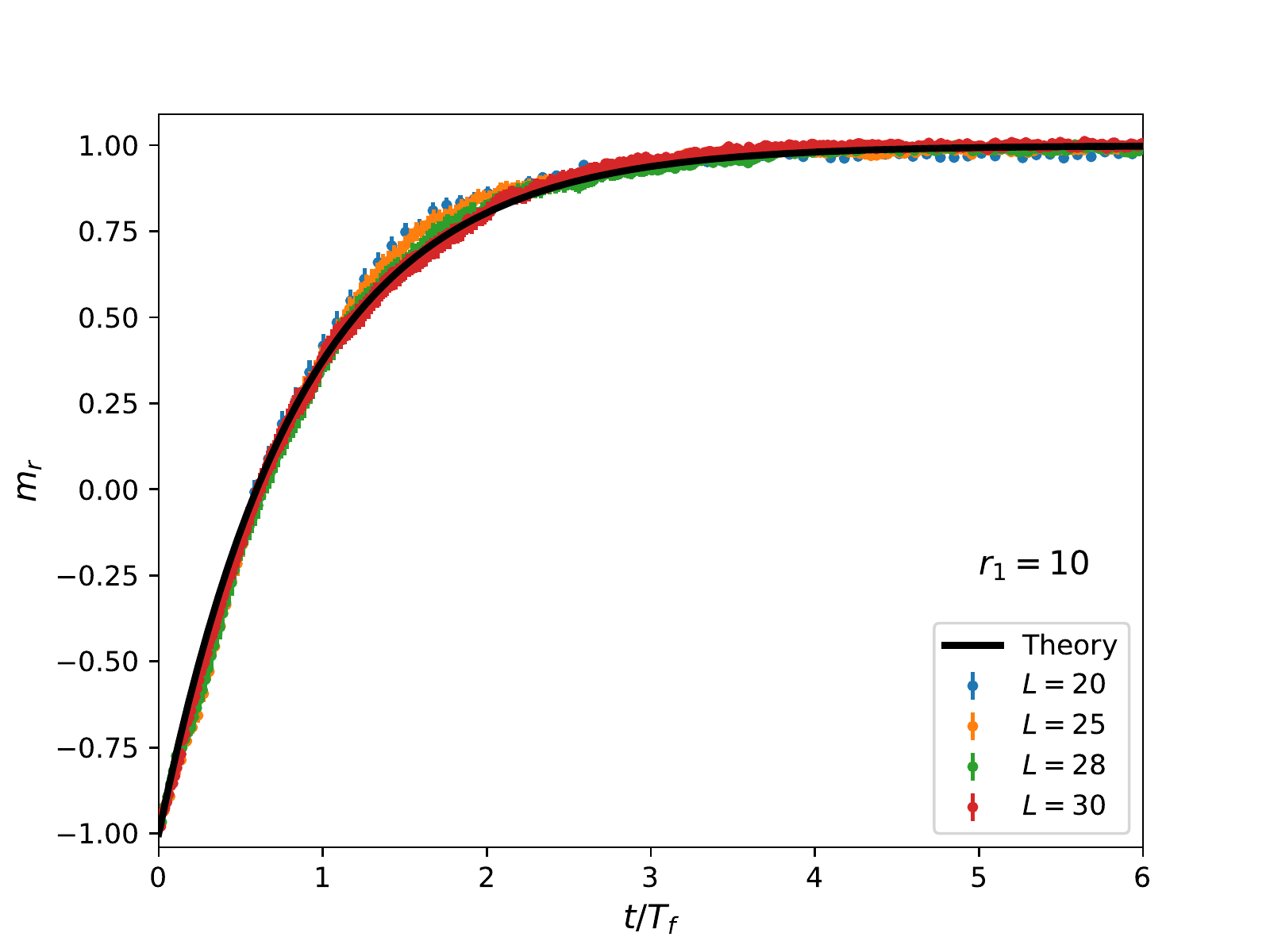}
	\caption{Renormalized magnetization $m_r=M/\tilde{m}_0$ as a function of $t/T_f(0.9,r_1,L)$ for $r_1=10$.}
	\label{OBC_mr10}
\end{figure}

Finally, we analyze the exponential nature of the renormalized magnetization. We use as time scale the estimated one for $\mu=0.9$ and consider the previous three values of $r_1$, i.e. $r_1=5,\;7.5,\;10$. The observed scaling behavior, reported in the figs. \ref{OBC_mr5}, \ref{OBC_mr75}, \ref{OBC_mr10}, is quite good, since the data collapse on a single curve. This confirms the exponential behavior of $m_r$ given by eq. \eqref{renormalized_magnetization_OBC} (black lines in the figures), with typical time scale given by $T_f(0.9,r_1,L)$.

\subsection{\label{z_exponent}Opposite boundaries: the equilibrium dynamic exponent $z$}
The equilibrium dynamic exponent $z$ of the relaxational dynamics is strictly related to the equilibrium large-$L$ behavior of the autocorrelation time $\tau$ of observables near the transition, i.e. $\tau\sim L^z$. We expect that at FOTs it depends on the boundary conditions. The case of PBC has already been analyzed \cite{berg_hansmann_neuhaus}, in particular $\tau$ is expected to exponentially increase with increasing $L$, i.e. $\tau\sim e^{\sigma L}$ where $\sigma$ is the interfacial free energy density and $z\rightarrow\infty$: this is related to the exponential increasing of the tunneling time between the coexisting phases at FOTs.

Here we consider OFBC: they are symmetric boundary coditions, as well as PBC, but promote explicitly the formation of an interface in the lattice. We thus expect that the behavior of the autocorrelation time drastically changes in this case: the dynamics near the transition is related to this interface, which moves within the lattice. This gives rise to a scaling law which is not an exponential but a power-law, i.e.
\begin{equation}
\tau\sim L^z,
\end{equation}
defining the equilibrium dynamic exponent of the Metropolis dynamics.

This behavior is analyzed for temperature-driven FOTs \cite{panagopoulos_vicari}, in particular for the 2D Potts model at $T=T_c$ with $q=20$ and mixed boundary conditions: in this case the
boundaries favor on one side the low-$T$ phase and at the opposite side the high-$T$ phase. Numerical results support such a power-law behavior for the integrated autocorrelation time. However we can not extend a priori this result to our case, since the type of transition is different (it is a field-driven FOT).

We numerically estimate $z$ by equilibrium MC simulations for various linear dimension $L\in[15,100]$. Details on the estimators used to estimate the autocorrelation time of the magnetization are reported in Appendix \ref{binning method}.

\begin{figure}[h]
	\centering
	\includegraphics[width=0.8\linewidth]{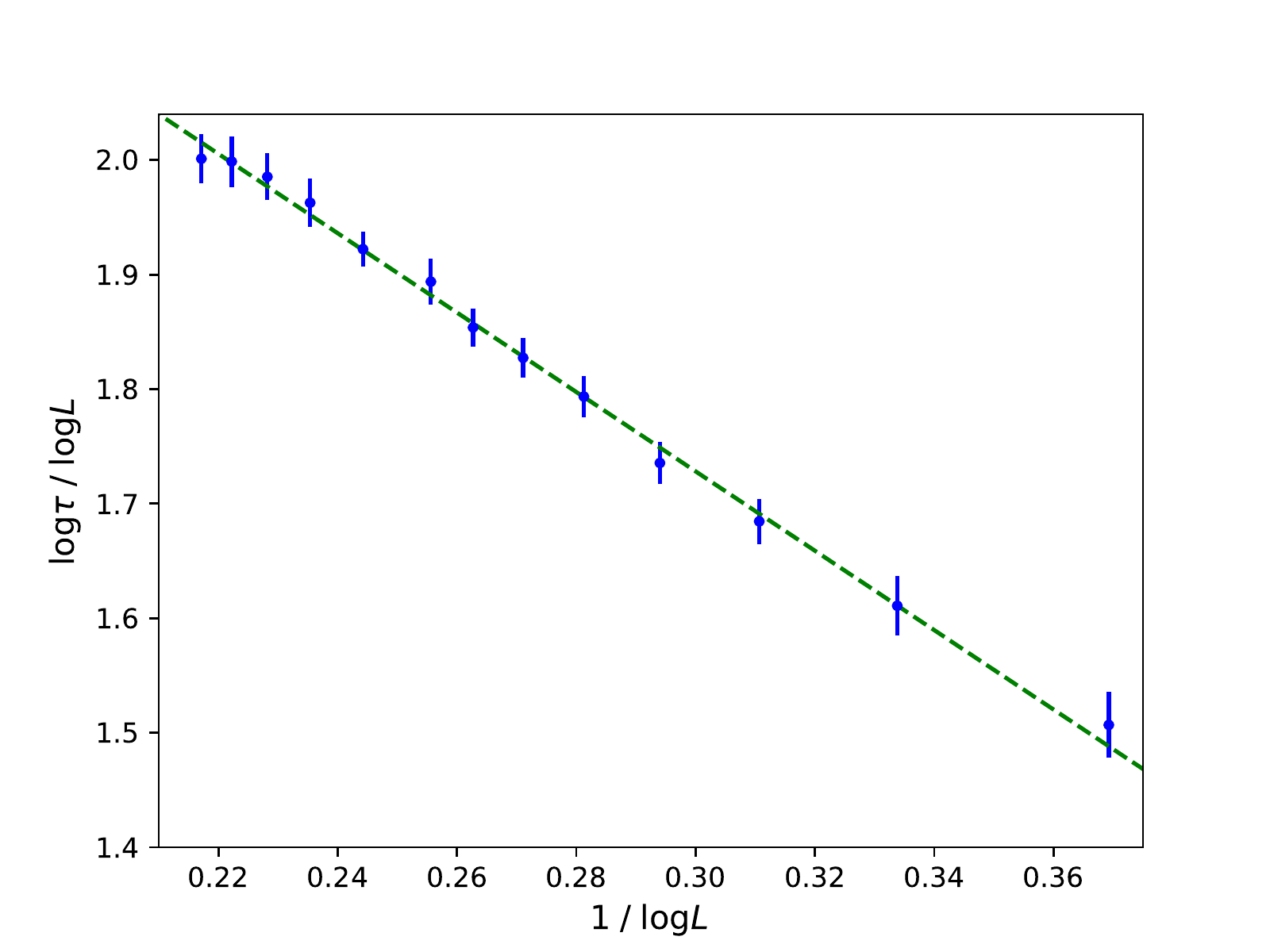}
	\caption{Log plot of the integrated autocorrelation time $\tau$ of the magnetization as a function of $1/\log{L}$. The dashed line is the linear fit with the value of $z$ reported in the first line of Table \ref{stime_z}.}
	\label{data_OFBC}
\end{figure}

In order to obtain an estimate of $z$ we make a log-log fit of the form
\begin{equation}
\log{\tau}=z\log{L}+q,
\end{equation}
since the log-log plot of the data is a straight line, as shown in fig. \ref{data_OFBC}.

In order to obtain a better estimate of the exponent we try to keep only some values of $L$, choosing only data with $L>L_{min}$. We observe that the values of the exponent obtained in this way are compatible within the errors. The results of the fit are reported in Table \ref{stime_z}.
\begin{table}[h]
	\setlength{\tabcolsep}{11pt}
	\begin{center}
		\begin{tabular}{ccc}
			\hline 
			\hline
			$L_{min}$ & $z$ & $\chi^2/\mathrm{ndof}$\\ 
			\hline 
			15 & $2.77(2)$ & $2.17/11$\\
			20 & $2.79(2)$ & $1.49/10$\\ 
			30 & $2.80(3)$ & $1.39/8$\\ 
			40 & $2.77(4)$ & $1.03/6$\\ 
			50 & $2.70(6)$ & $0.58/4$\\ 
			\hline
			\hline
		\end{tabular}
	\end{center}
	\caption{Values of $z$ for different choices of $L_{min}$. $\chi^2/\mathrm{ndof}\sim0.2$ for all values of $L_{min}$, being ndof the number of the degrees of freedom of the fit.}
	\label{stime_z}
\end{table}

Given the values of $z$ in Table \ref{stime_z} we can conclude by considering
\begin{equation}
z=2.77(4)
\end{equation}
as our final estimate of the equilibrium dynamic exponent in the case of OFBC on a square 2D Ising lattice.

It is necessary to make some general comments on the value of $z$: its value should be intrinsically related to the interface dynamics within the lattice, thus we could try to extend the obtained result to other finite systems at magnetic FOTs with boundary conditions that favor the formation of such an interface. Furthermore, it should extend to the whole class of purely relaxational dynamics, including also the heat bath upgrading. As a consequence, different classes of dynamics may lead to other values of the dynamic exponent $z$.

\section{\label{conclusions} Conclusions}
We have studied how the scaling properties of finite systems near a FOT are affected by a change in the boundary conditions. In particular, we have analyzed the behavior of Ising systems, showing a field-driven FOT, in the case of a purely relaxational dynamics.

We have first considered the case of OFBC, i.e. generating an interface in the system. We have analyzed the equilibrium FSS of the average magnetization, showing that it scales with the variable $r_1=hL^2$, as it happens for the PBC case. However, this scaling behavior can be related to the movement of the interface within the lattice. 

We have then analyzed the dynamic scaling behavior in the coexistence region, showing that the corresponding DFSS theory developed for PBC can be extended to the case of finite systems with OBC. In this region the system oscillates among the two different coexisting phases at the transition, and the relevant time scale of the dynamics is the tunneling time between the two phases, which scales as $\tau(L)\approx L^\alpha e^{\kappa\beta L}$. On time scales of the order of $\tau$ the dynamic observed through MC simulations seems to confirm our conjectures about the evolution of the magnetization. Data obtained for the mean FPT allowed us to evaluate the scaling behavior of $\tau$, confirming in particular that the dynamic behavior in the coexistence region is ruled by the formation of a planar interface in the lattice.
Then we have observed the finite-size behavior of the integrated autocorrelation time of the magnetization in the OFBC case: we have found a power-law behavior $\tau\sim L^z$, where $z\approx2.8$. This behavior is different compared to the PBC case, where $\tau$ is exponential in $L$, and we argued that this may be related to the dynamics of the interface generated by the boundary conditions.

All these analyses show that the finite-size behavior of systems close to a FOT depends effectively on the boundary conditions considered. In particular, the understanding of these effects for general boundaries, i.e. not necessarily periodic, is of physical interest: conditions that generate interfaces separating the phases, like the OFBC, are of experimental interest, for instance in experiments of small systems, such as those considered for our tests, when the characteristic time scales of the system are of the order of the time scale of the experiment. 

Other types of interesting boundaries that could be considered are those that favor one of the phases of the system, like equally fixed boundary conditions (EFBC) in the Ising case. Indeed, in a recent study concerning the FSS properties of the quantum Ising chain at quantum FOTs \cite{pelissetto_rossini_vicari}, a new interesting equilibrium scaling behavior emerges when considering EFBC in the region characterized by the physical coexistence of the phases in the system. This turns out to be governed by the scaling variable $r_1=hL^\varepsilon$, where $\varepsilon=2$ (a value that can be explained using theoretical arguments, also supported by numerical results). Using arguments related to the quantum-to-classical mapping, we expect that an analogous phenomenon should be observed at FOTs in the 2D Ising model defined in slab geometries. This type of scaling is also observed in the case of thermal FOTs \cite{Panagopoulos_pelissetto_vicari_anomalousfss}, e.g. considering the 2D Potts model with $q>4$ states and OBC, and turns out to be relevant for heavy-ion experiments searching for evidence of FOTs in the hadron phase diagram \cite{PETERSEN2017145}.

We finally mention that, in the view of potential future developments, it could also be interesting to study the off-equilibrium behavior of Ising systems at FOTs. One may indeed consider an off-equilibrium dynamics driven by a time-dependent magnetic field in order to see if a nontrivial scaling behavior is observed when the transition point is slowly crossed, analogously to what happens in the case of thermal FOTs of the 2D Potts model \cite{pelissetto_vicari_offequilibrium,Panagopoulos_pelissetto_vicari_anomalousfss}.

\section*{\label{ringraziamenti}Acknowledgements}
The author thanks sincerely Ettore Vicari, for useful discussions and for all the help provided during the drafting of this paper.

\appendix 
\section{\label{binning method}Computation of $\tau$}
We could estimate the integrated autocorrelation time of the magnetization by means of its definition
\begin{equation}
\tau=\frac{1}{2}\sum_{t=-\infty}^{t=+\infty}\frac{\langle(M(t)-\langle M\rangle)(M(0)-\langle M\rangle)\rangle}{\langle(M(0)-\langle M\rangle)^2\rangle}    
\end{equation}
where averages are taken at the equilibrium. Instead of working with the autocorrelation functions, we use an estimator of $\tau$ that can be obtained by the binning method \cite{WOLFF2004143,DELDEBBIO2004315}
\begin{equation}
\tau=\frac{E^2}{2E_0^2},
\end{equation}
where $E$ is the error found after binning, i.e. when the error computed with this method becomes stable with respect to an increase of the block size $b$, and $E_0$ is the error computed using all the measures for $M(t)$ directly. Clearly $E_0$ does not take into account the autocorrelation between measures.

Denoting with $N_b$ the number of blocks used in the computation of $E$, the statistical error $\Delta\tau$ associated to $\tau$ is
\begin{equation}
\Delta\tau=\sqrt{\frac{2}{N_b}}\tau.
\end{equation}
The procedure leads to a systematic error $\delta\tau=O(\tau/b)$ which is negligible if $\delta\tau\ll\Delta\tau$. In our estimates of Section \ref{z_exponent} this approximation is verified since $\delta\tau$ is always $O(10^{-3})$, or $O(10^{-2})$ for $L\geq60$, while statistical error varies in a range from $\Delta\tau=O(10)$ to $\Delta\tau=O(10^3)$.

\end{document}